\documentclass[10pt,journal,twoside,web]{ieeecolor}
\usepackage{generic}
\usepackage{cite}
\usepackage{amsmath,amssymb,bm,mathtools}
\usepackage{algorithmic}
\usepackage{graphicx,subcaption}
\usepackage{textcomp}
\usepackage{comment}
\usepackage{algorithm,algorithmic}
\usepackage{graphicx}
\usepackage{xcolor}
\let\labelindent\relax
\usepackage{enumitem}
\usepackage{lipsum}
\usepackage{ntheorem}
\usepackage[normalem]{ulem}

\newcommand{\Initialize}[1][1]{
  \STATE \textbf{Initialize:}
}
\newtheorem{theorem}{Theorem}

\newtheorem{lemma}[theorem]{Lemma}

\newtheorem{assumption}{Assumption}
\newtheorem{definition}{Definition}
\newcommand{\call}{\mathcal{L}}
\newcommand{\caln}{\mathcal{N}}
\newcommand{\calv}{\mathcal{V}}

\newcommand{\bo}[1]{{\textcolor{blue}{Bo says: #1}}}

\def\BibTeX{{\rm B\kern-.05em{\sc i\kern-.025em b}\kern-.08em
    T\kern-.1667em\lower.7ex\hbox{E}\kern-.125emX}}
\begin{document}

\title{Online Network Utility Maximization: Algorithm, Competitive Analysis, and Applications}
\author{Ying~Cao, Bo~Sun, and Danny~H.K.~Tsang, \IEEEmembership{IEEE Fellow}
% \thanks{This work was supported by .}
\thanks{The authors are with Department of Electrical and Computer Engineering, Hong Kong University of Science and Technology. (e-mail: ycaoan, bsunaa, eetsang@ust.hk). }
}

\maketitle

\begin{abstract}
We consider an online version of the well-studied network utility maximization problem, where users arrive one by one and an operator makes irrevocable decisions for each user without knowing the details of future arrivals. We propose a threshold-based algorithm and analyze its worst-case performance. We prove that the competitive ratio of the proposed algorithm is linearly increasing in the number of links in a network and show this competitive analysis is tight. Extensive trace-driven simulations are conducted to demonstrate the performance of our proposed algorithm. In addition, since worst-case scenarios rarely occur in practice, we devise an adaptive implementation of our algorithm to improve its average-case performance and validate its effectiveness via simulations.
\end{abstract}

\section{Introduction} \label{sec:introduction}

\IEEEPARstart{N}{etwork} utility maximization (NUM) is a general optimization paradigm of vital importance in the field of networking. It has been widely applied after the seminal work by Kelly \textit{et al.}~\cite{kelly1998rate}. For example, it often serves as the underlying model to draw insights on understanding and designing congestion control mechanisms in computer networks~\cite{kelly1998rate} and media access control protocols in wireless networks~\cite{congestionControlWireless,lee2006optimal}. In addition, the utility-based models are shown to be effective for the power-aware load balancing and queueing in the communication network management~\cite{utilityBasedNM}. Utility-based maximization has also been shown to play an important role in areas outside traditional communication networks, such as the demand response in power systems~\cite{li2011optimal}, the electric vehicle charging control~\cite{evcharging}, information dissemination in the vehicular ad-hoc networks~\cite{dataDisVanet} and many other applications~\cite{videoStreaming,dsl,sensorNetwork,chen2010utility}.

Existing research on NUM usually assumes users are static. Therefore, they mainly focus on designing distributed algorithms to solve the offline problem. In reality, users usually arrive one by one in an online manner. For example, hosts come sequentially requesting network access in public networks. In addition, cloud service requests arrive one by one to a cloud data center. Thus, we consider an online version of NUM and term it as online network utility maximization (ONUM). Compared with its offline counterpart, the difficulty of ONUM originates from the fact that the information about the problem, e.g., the utility function, is revealed piece by piece. We need to make irrecoverable decisions that only depend on the causal information (i.e., the past and current information), and the aim is to be as close as possible to the offline optimum that can be obtained if all information is given from the start. 

There exist some efforts in the literature that deal with the ONUM problem. Some works assume that utility functions are drawn i.i.d. from an unknown distribution and use online learning to solve ONUM under the regret minimization framework, such as~\cite{lu2020dual,agrawal2014fast}. However, the i.i.d. assumption may not be valid in reality, and it is worth noting that the capacity constraints will be violated in those algorithms. Another line of work assumes that the utility functions are generated adversarially and designs algorithms under the competitive analysis framework. For instance, \cite{buchbinder2009online} considers the online packing problem and \cite{agrawal2014dynamic} considers the online linear programming problem. However, both of them allow violation of the constraints. One special case in \cite{buchbinder2009online} can ensure no constraints are violated when all constraints coefficients are either 0 or 1; however, they just consider a simple linear objective without uncertainty. \cite{convex2016} and \cite{huang2019welfare} consider non-linear uncertain objectives but the non-linearity source is different from ours, which will be clear in the next section. In this paper, we consider designing online algorithms for the general ONUM under the competitive analysis framework and can ensure no violation of the constraints.

A systematic way to design and analyze competitive online algorithms is to follow the online primal-dual framework, which is based on the weak duality and has been applied successfully to several classic online problems, such as online covering and packing\cite{buchbinder2009online}\cite{buchbinder2007online}, online matching\cite{huang2020online} and weighted paging\cite{bansal2012primal}, etc. The algorithm proposed in this paper is also related to the dual, however, the dual problem for the ONUM problem with general concave utility functions is not able to be expressed as explicitly as the counterpart problems mentioned before. Thus, we bypass the online primal-dual analysis method in this paper and adopt a different analysis method, which directly bounds the competitive ratio by identifying the worst-case instances.

The contributions of this paper consist of three parts:
\begin{itemize}
    \item \textbf{Algorithm.} We design an online \textit{threshold-based algorithm} for ONUM with general concave utility functions and hard capacity constraints. The threshold is an increasing function of the resource utilization level, whose curvature tunes the behavior of the algorithm, balancing between being aggressive and conservative.
    \item \textbf{Competitive Analysis.} We show under the umbrella of the competitive analysis framework, that when the threshold function satisfies certain \textit{sufficient conditions}, the algorithm yields a bounded competitive ratio. We adopt a direct analysis method which does not rely on the dual and prove the tightness of the analysis by characterizing the worst-case instances.
    \item \textbf{Applications.} We apply the algorithm to real network traces and propose a practical adaptive implementation based on \textit{online learning} that can tune the parameters of the threshold-based algorithm, yielding much better empirical results.
\end{itemize}

This paper is organized as follows. In Section \ref{sec:onum}, we present the system model for ONUM and show some applications that fit the model. In Section \ref{sec:alg}, we present the threshold-based algorithm, explain the intuition, and give a tight competitive analysis of the algorithm rigorously. In Section \ref{sec:simulation}, we first conduct simulations to demonstrate the proposed algorithm's performance, compare it with two heuristic algorithms under various arrival patterns, and propose viable adaptive implementation methods that strategically adjust the algorithm parameters to further improve the empirical results. In Section \ref{sec:conclusions}, we conclude the paper and discuss about promising future directions.

\section{Online Network Utility Maximization} \label{sec:onum}
% ONUM settings and assumptions
We describe the general system model for ONUM and show some exemplary applications in this section.
A network with a link set $\mathcal{L}$ is considered, where each link $\ell\in\call$ has a capacity of $c_\ell$. Let $L$ be the total number of links. We consider the homogeneous capacity case, i.e., $c_\ell=1$ for any link.
We let a user represent a traffic flow with a source and a destination. Users come to the network one by one. The transmission rate of the user needs to be determined upon its arrival and the allocation is irrevocable. The routing path of a user is a set of links that connect its source and destination. For the $i$th user, denote its routing path as $\mathcal{L}_i$. In our problem, we assume that the routing path is determined before the arrival of the user. The capacity consumption at each link on the routing path of a user is equal to its rate allocation. The total rate allocation on any link cannot exceed its capacity. In addition, the user also comes with a utility function $g_i(\cdot)$ and a budget $b_i$. The utility function is a function of the rate allocated to the user and the budget is the highest rate at which the user is able to transmit. After allocating an amount of rate $y_i$ to the $i$th user, $g_i(y_i)$ amount of utility can be gained. We denote user $i$ by the tuple $A_i =\{g_i(\cdot), \mathcal{L}_i, b_i\}$.
Let $\caln$ denote the set of users and let $N$ be the total number of users.
The goal is to design an online algorithm that determines the rate allocation $y_i$ by the time of $i$th arrival, maximizing the total utility of all users. Note that $y_i$ is determined without knowing the utility functions of future users, i.e., $\{g_k(\cdot)\}_{k>i}$. If the full sequence of arrivals $\mathcal{I}=\{A_1,\dots,A_N\}$ is disclosed at the beginning, our problem is formulated as follows:
\begin{subequations}
\label{offline-num}
\begin{align}
\max_{y_i} \quad &\sum_{i\in\caln} g_i(y_i) \\
{\rm s.t.} \quad &\sum_{i:\ell\in\mathcal{L}_i}y_i\le 1, \forall \ell\in\mathcal{L},\\
& 0\le y_i \le b_i,\forall i\in\caln .
\end{align}
\end{subequations}

A convex program solver can solve Problem (\ref{offline-num}) optimally. We denote the optimal objective as OPT($\mathcal{I}$). The performance of an online algorithm is usually evaluated by the \textit{competitive analysis}\cite{borodin2005online}. Given a request sequence $\mathcal{I}$, let ALG($\mathcal{I}$) be the objective value achieved by an online algorithm. The performance of the online algorithm is evaluated by the competitive ratio defined as 
\begin{align*}
    \alpha=\max_{\forall \mathcal{I}} \frac{\text{OPT}(\mathcal{I})}{\text{ALG}(\mathcal{I})}.
\end{align*}
The competitive ratio is desired to be as small as possible, such that the algorithm performs as close as possible to the offline optimum even under the worst case.

We make the following assumptions on the utility functions.
\begin{assumption}\label{assumption-1}
    $g_i(y)$ is increasing, strictly concave and continuously differentiable over $[0,b_i]$ and $g_i(0)=0$.
\end{assumption}
\begin{assumption}\label{assumption-2}
    the marginal utility averaged over the number of links is bounded from above and below, i.e., $\frac{g'_i(y)}{|\mathcal{L}_i|}\in [m,M]$.
\end{assumption}
Assumption \ref{assumption-1} is standard for NUM problems~(e.g., \cite{kelly1998rate,lee2006optimal,li2011optimal}). It means that the user utility is usually increasing in the rate allocated and is concave because of the diminishing returns property. The commonly used utility functions are logarithm functions and polynomial functions, and thus the differentiability assumption is usually satisfied. Assumption \ref{assumption-2} requires the first-order derivatives of utility functions to be bounded, which is essential to achieve a bounded competitive ratio. Similar assumptions appear in other online optimization literature, e.g., online knapsack problem~\cite{zhou2008budget,zhang2017optimal} and one-way trading problem~\cite{el2001optimal}.

%This implies that $g_i(y)$ is linearly increasing with the number of links user $i$ requests, which is true for some applications, such as information dissemination networks. For other applications such as rate control, another assumption that $g'_i(y)\in[m,M]$ is more appropriate, and we are also able to prove the same competitive results, with a minor modification in the algorithm (see Appendix). However, the current assumptions allow for a clearer presentation of analysis.

We now provide some exemplary applications that fit into the ONUM model.
% \begin{itemize}

\textbf{Online routing of virtual circuits~\cite{awerbuch1993throughput}.}
In the simplest version, requests $r_i=(s_i,t_i)$ come online with a predetermined routing path between source $s_i$ and destination $t_i$. The algorithm will determine whether or not the request can be accepted. If the request is accepted, the algorithm then decides how much bandwidth $y_i$ should be allocated to the request and establish a virtual circuit with the requested routing path. The aggregate throughput is $\sum_{i}y_i$, and thus a throughput-maximizing objective is linear in the allocated bandwidth. The methods and results developed in this paper can be easily applied to this case.

\textbf{Online flow control for wireless sensor networks~\cite{chen2010utility}.}
A sensor network is modeled as a connected graph $G(\calv,\call)$, where $\calv$ denotes the sensor nodes and $\call$ denotes the logical bidirectional communication links between the sensor nodes. Due to the broadcast nature of sensors, each link $\ell\in \call$ has an interference link set $IS_\ell$. Each sensor node $v\in \calv$ has an energy capacity $e_v$ and each link $\ell \in \call$ has an interference margin level $c_\ell$, which guarantees the transmission rate of the flow on a link if the margin level is observed by all flows in the link's interference link set. Traffic flows are generated online. For the $i$th flow, it goes through a set of sensors $\calv_i$ and a set of links $\call_i$ with the transmission rate $y_i$ to be determined. Also, the $i$th flow is characterized by a utility function $g_i(y_i)$ that is strictly concave in $y_i$. There are two sets of constraints, the link capacity constraints $\sum_{\ell'\in IS_\ell}\sum_{i: \ell'\in \call_i}y_i \le c_\ell$ for each link $\ell\in \call$ and the energy constraints $(e^t+e^r)\sum_{i: v\in \calv_i}y_i \le e_v/T-e^d$ for each sensor $v\in \calv$, where $T$ is the pre-specified sensor lifetime, $e^t, e^r$ and $e^d$ are the energy consumption per unit data during the transmission, reception and idle state, respectively. The goal is to maximize the sum of utility $\sum_{i}g_i(y_i)$ subject to the two sets of constraints. It can be studied under the ONUM model by simply normalizing the parameters.

\textbf{Online rate control of elastic traffic~\cite{la2002utility}.}
Consider a network with a set $\call$ of resources, and let $c_\ell$ be the finite capacity of resource $
\ell$. Users arrive online to generate traffic along its chosen route, which is a subset of $\call$. The same route can be taken by multiple users. Let the route chosen by the $i$th user be $\call_i$. Suppose that if rate $y_i$ is allocated to the traffic of the $i$th user, a utility of $g_i(y_i)$ is gained by the user. The traffic is called elastic when the utility function is increasing, concave and continuously differentiable function of $y_i$ over $y_i\ge 0$. The problem is to find the optimal rate allocation for sequentially-arriving elastic traffic users to maximize the aggregate utility of rate $\sum_{i}g_i(y_i)$ subject to the resource capacity constraints.
% \end{itemize}

\begin{comment}
\begin{definition}[Arrival Instance]
An arrival instance $\mathcal{I}$ is a fixed sequence of utility functions $\{g_i(y)\}_{t\in \mathcal{T}}$.
\end{definition}

When $\mathcal{I}$ is known beforehand, the problem is equivalent to the well-studied network utility maximization problem (NUM)\cite{palomar2006tutorial}, and can be stated as
\begin{align}
    \label{basic-num}
    \max_{x_i} \quad &\sum_{i=1}^T g_i(y_t) \\
    \nonumber
    s.t. \quad &\sum_{t:\ell\in \mathcal{L}_i} y_t\le 1, \forall \ell \in \mathcal{L} \\
    \nonumber
    &0\le y_t \le b_i, \forall t\in \mathcal{T}.
\end{align}
In comparison, $\{g_i(\cdot)\}$ appears online in our problem.
\end{comment}

\section{Competitive Online Algorithms for ONUM} \label{sec:alg}
% Algorithm and its analysis
In this section, we present an online threshold-based algorithm for ONUM, establish sufficient conditions for the algorithm to have a bounded competitive ratio, and show the tightness of the competitive analysis.
% of the theorem based on the online primal-dual analysis framework.
Before introducing the algorithm, we first introduce an idea that is not uncommon for algorithms that rely on the dual problems.
In the algorithmic design based on an optimization problem, dual variables are usually viewed as the prices, reflecting the system state. One standard distributed algorithm to solve the classic NUM problem \cite{dual-update} is based on this idea, where each link is attributed a dual price $\lambda_\ell$ and user $i$ determines its rate $y_i$ by solving the following maximization problem:
\begin{align}\label{eq1}
    y_i^*(\lambda^i) = \arg\max_{y_i\ge0} \left(g_i(y_i)-\lambda^i y_i\right), \forall i,
\end{align}
where $\lambda^i=\sum_{i:\ell\in\call_i}\lambda_\ell$ is the total price on the path of user $i$. 
The dual prices of links will be updated as follows:
\begin{align}\label{eq2}
    \lambda_\ell(t+1) = \left[ \lambda_\ell(t)-\gamma\left(c_\ell-\sum_{i:\ell\in\call_i}y_i^*(\lambda^i(t))\right) \right]^+, \forall \ell,
\end{align}
where $t$ is the iteration index and $\gamma$ is the step size. By iterating between \eqref{eq1} and \eqref{eq2}, $y_i^*$ and $\lambda_\ell$ converge to the optimal rate allocation and dual prices \cite{dual-update}.
In the design of online algorithms, we also aim to determine the online rate allocation based on equation~\eqref{eq1}. However, in the online setting, the rate allocation is irrevocable, making it impossible to iteratively update the prices based on equation~\eqref{eq2}. Thus, the key is to design an approach to update prices in an online manner. 
In our online algorithm, we design the link dual price as a function of the link utilization level and introduce the following definition:
\begin{definition}[Value Function]
A value function for link $\ell$, $\phi_\ell(\omega_\ell): [0,1]\rightarrow \mathbb{R}^+$, is a non-decreasing continuous function that evaluates the marginal utility of the resource at the utilization level $\omega_\ell$.
\end{definition}
Contrary to equation~\eqref{eq2}, where the dual price of each link is determined based on the total utilization of the link, the value function provides a way of estimating the link price based on the currently observed utilization level. In general, our algorithm takes the value function $\phi:=\{\phi_\ell\}_{\ell\in\call}$ as input and sequentially determines the rate allocation $y_i$ for each user upon observing its request tuple $A_i$.
Next, we present our proposed algorithm in Algorithm~\ref{alg:oa} and term it as the online algorithm with value function $\phi$ ($\text{OA}_\phi$).
In detail, $\text{OA}_\phi$ uses the value function $\phi_\ell(\cdot)$ to evaluate the scarcity of the remaining capacity of link $\ell$ by the link price.
We denote by $\omega_\ell^i$ the total capacity of link $\ell$ consumed by the previous $i$ users. 
The cost of using an infinitesimal amount of capacity, $ds$, of link $\ell$ when its utilization is $s$ can be estimated by $\phi_\ell(s)ds$, and thus the cost of using link $\ell$ by user $i$ can be shown as $\int_{w_\ell^{i-1}}^{w_\ell^{i-1}+y_i}\phi_\ell(s)ds$.
$\text{OA}_\phi$ then determines $y_i$ by solving a pseudo-utility maximization problem~\eqref{decision-rule}, where the user's pseudo-utility is defined as its utility of being allocated rate $y_i$, i.e., $g_i(y_i)$, minus the total cost of using links.
Since all the design space of $\text{OA}_\phi$ lies in the value function, the following part of this section will be centered around the key question: what conditions should $\phi$ follow such that $\text{OA}_\phi$ can yield a bounded competitive ratio?
\begin{algorithm}
\caption{Online Algorithm with Value Functions $\phi$ ($\text{OA}_\phi$)}
\begin{algorithmic}\label{alg:oa}
\Initialize value function $\phi$, initial utilization $\omega^0_\ell = 0, \forall \ell$;
\FOR{the $i$th user}
\STATE Observe user $i$'s request $A_i =\{g_i(\cdot), \mathcal{L}_i, b_i\}$;
\STATE Determine $y_i$ by solving the  problem
\begin{align}
\label{decision-rule}
    y_i= \arg\max_{0\le y\le b_i}  g_i(y) - \sum_{\ell\in \mathcal{L}_i}\int_{\omega_\ell^{i-1}}^{\omega_\ell^{i-1}+y} \phi_\ell(s)ds;
\end{align}
\STATE Update for links $\ell \in \mathcal{L}_i$: $\omega_\ell^i =  \omega_\ell^{i-1}+y_i;$
\ENDFOR
\end{algorithmic}
\end{algorithm}

\subsection{Sufficient Conditions for Being Competitive}
The following theorem provides a sufficient condition on the value function $\phi$ to ensure $\text{OA}_\phi$ to be competitive.

\begin{theorem}\label{thm: suff_conditions}
$\text{OA}_\phi$ is $L\alpha$-competitive if $\phi_\ell$ is given by
\begin{align}
    \label{suffi_conditions}
    \phi_\ell(y) = \begin{cases}
    m, & y \in [0,\beta),\\
    \varphi_\ell(y), & y \in [\beta,1],\\
    \infty, & y\in (1,\infty),
\end{cases}
\end{align}
where 
\begin{enumerate}[label=(\roman*)]
    \item $\alpha\ge \ln(M/m)+1$. $\beta\in[0,1]$ is a utilization threshold and satisfies
    \begin{align}
    \label{sufficient-condition-1}
        \beta \ge 1/\alpha,
    \end{align}
    \item $\varphi_\ell$ is a non-decreasing function that satisfies
\begin{align}
\label{sufficient-condition-2}
    \begin{cases}
        \varphi_\ell(y) \ge \frac{1}{\alpha} {\varphi_\ell}'(y),y \in [\beta,1], \\
        \varphi_\ell(\beta) = m, \varphi_\ell(1) = M.
    \end{cases}
\end{align}
\end{enumerate}
\end{theorem}
Before the detailed proof, we provide intuitions on the sufficient conditions above. We note that when all users request for the same single link $\ell$ with $g_i'(y)=m$, i.e., all users are with sufficiently low utility, if $\phi_{\ell}(0)>m$, all of them will be rejected. To ensure a bounded competitive ratio for this case, the value function needs a flat segment with $\phi_\ell(y)=m$ at the beginning. After that, when the remaining resource decreases, the link price increases so as to preserve capacity for possible high-utility users. The exponential form is the same as the single link case in our previous work~\cite{cao2020optimal}. We next show the proof of Theorem~\ref{thm: suff_conditions}.

\begin{proof}
% We use the online primal-dual analysis to prove Thm \ref{thm: suff_conditions}. Since for any link $\ell$, $\phi_\ell(\omega)=\infty,\omega >c_\ell$, the online algorithm will always produce a feasible set of decisions $\{y_i\}$ that does not violate the capacity constraints, i.e., $\sum_{t:\ell \in \mathcal{L}_i} y_i \le c_\ell,\forall \ell$.
Because $\text{OA}_\phi$ is deterministic, for any arrival instance $\mathcal{I}$, $\text{OA}_\phi$ will produce a deterministic final utilization level for each link. Also, different arrival instances can induce the same set of final link utilization levels. Denote the final utilization level of link $\ell$ by $\omega_\ell^N$, i.e., $\omega^N_\ell=\sum_{i:\ell\in \mathcal{L}_i}y_i$. We are then able to classify the set of all possible arrival instances $\mathcal{I}$ into 3 cases according to their induced final utilization levels:
\begin{itemize}
    \item \textbf{Case 1}: $\forall \ell \in \mathcal{L}, \omega_\ell^N < \beta$.
    \item \textbf{Case 2}: $\forall \ell\in\mathcal{L},\omega_\ell^N<1$; $\exists \ell\in\mathcal{L}, \omega_\ell^N \ge \beta$.
    \item \textbf{Case 3}: $\exists \ell\in\mathcal{L}, \omega_\ell^N=1$.
\end{itemize}

In Case 1, we make the observation that any user will transmit at its maximal rate $b_i$.
Because for any arrival $i$, $\omega_\ell^i \le \omega_\ell^N < \beta$, $\phi(\omega_\ell^i)=m\le g'_i(b_i)\le g'_i(y)$, then $ g_i(y) - \sum_{\ell\in \mathcal{L}_i}\int_{\omega_\ell^{i-1}}^{\omega_\ell^{i-1}+y} \phi_\ell(s)ds$ is non-decreasing in $y$ over $[0,b_i]$ and $y_i=b_i$ is the maximizing solution. In this case, the offline optimal decisions are exactly the same as those made by the online algorithm, and thus the offline optimal utility is equal to the online utility, i.e., OPT($\mathcal{I}$) = ALG($\mathcal{I}$)$ \le L\alpha$ALG($\mathcal{I}$), where both the number of links $L$ and parameter $\alpha$ are no less than 1.

In Case 2, there exists a link whose final utilization level exceeds $\beta$, which makes the exponential segment take effect. Moreover, no link has reached the capacity limit in Case 2. Given the final utilization $\{\omega_\ell^N\}_{\ell\in\call}$, we next find the worst case of $\mathcal{I}$, for which the ratio $\frac{\text{OPT}(\mathcal{I})}{\text{ALG}(\mathcal{I})}$ is maximized over all possible instances in Case 2.

For an arrival instance $\mathcal{I}$, denote the last user with $y_i>0$ by $\tau$. We group the users that come before $\tau$ into 4 sets, based on their utility functions and requested link set $\{g_i(\cdot),\mathcal{L}_i\}$. Let $\caln_h$ denote the set of \textit{high-valuation} users whose minimal marginal utility $g'_i(b_i)\ge\sum_{\ell\in \mathcal{L}_i}\phi_\ell(\omega_\ell^N)$, and thus for $i \in \caln_h$, $y_i=b_i$; Let $\caln_m$ denote the set of \textit{medium-valuation} users that $\exists y \in (0,b_i)$ such that $g'_i(y)=\sum_{\ell\in \mathcal{L}_i}\phi_\ell(\omega_\ell^N)$, and thus for $i\in\caln_m$, $y_i>0$; Let $\caln_l$ denote the set of \textit{low-valuation} users whose largest marginal utility $g'_i(0)\le\sum_{\ell\in \mathcal{L}_i}\phi_\ell(\omega_\ell^N)$ and $g'_i(b_i)<\sum_{\ell\in \mathcal{L}_i}\phi_\ell(\omega_\ell^i)<g'_i(0)$, and thus for $i\in\caln_l$, $y_i>0$; Let $\caln_z$ denote the set of \textit{very-low-valuation} users that $g'_i(0)\le\sum_{\ell\in \mathcal{L}_i}\phi_\ell(\omega_\ell^i)\le\sum_{\ell\in \mathcal{L}_i}\phi_\ell(\omega_\ell^N)$, and thus $y_i=0$ for $i\in\caln_z$. Note that either of the aforementioned four user sets is an empty set if there do not exist such users in the arrival instance $\mathcal{I}$. We use $\Tilde{y}_i$ to denote the offline optimal rate allocation decision for user $i$, and define $\hat{y}_i$ as the value such that $g'_i(\hat{y}_i)=\sum_{\ell\in \mathcal{L}_i}\phi_\ell(\omega_\ell^N)$ for $i\in \caln_m$. The offline optimal total utility for any $\mathcal{I}$ is
    \begin{equation}
    \label{eq: opt_expression}
    \begin{split}
        \text{OPT}(\mathcal{I})&= \sum_{i=1}^\tau g_i(\Tilde{y}_i)+\sum_{i=\tau+1}^N g_i(\Tilde{y}_i) \\
        &\overset{(i)}{=} \sum_{i\in\caln_h} g_i(b_i)+\sum_{i\in \caln_m} g_i(\Tilde{y}_i)+\sum_{i\in \caln_l} g_i(\Tilde{y}_i) \\
        &+ \sum_{i\in \caln_z} g_i(\Tilde{y}_i) + \sum_{i=\tau+1}^N \int_{0}^{\Tilde{y}_i}g_i'(s)ds \\
        &\overset{(ii)}{\le} \sum_{i\in\caln_h} g_i(b_i)+\sum_{i\in \caln_m} \big[g_i(\hat{y}_i)+\sum_{\ell\in\mathcal{L}_i} \phi_\ell(\omega_\ell^N)(\Tilde{y}_i-\hat{y}_i)\big]  \\
        &+ \sum_{i\in \caln_l\cup\caln_z} \sum_{\ell\in\mathcal{L}_i} \phi_\ell(\omega_\ell^N)\Tilde{y}_i + \sum_{i=\tau+1}^N \int_{0}^{\Tilde{y}_i}g_i'(s)ds\\
        &\overset{(iii)}{\le} \sum_{i\in\caln_h} \big[g_i(b_i)-\sum_{\ell\in\mathcal{L}_i}\phi_\ell(\omega_\ell^N)b_i\big] \\
        &+ \sum_{i\in \caln_m} \left[g_i(\hat{y}_i)-\sum_{\ell\in\mathcal{L}_i}\phi_\ell(\omega_\ell^N)\hat{y}_i\right] +\sum_{\ell\in\mathcal{L}}\phi_\ell(\omega_\ell^N).
    \end{split}
    \end{equation}
% &\overset{(g)}{\le} \sum_{i\in\mathcal{T}_h} g_i(b_i)+\sum_{i\in \mathcal{T}_m} g_i(\hat{y}_i))+ \sum_{\ell\in\mathcal{L}}\phi_\ell(\omega_\ell^N)(1-\delta^\tau_\ell),
Equality $(i)$ holds due to the observations that, for high-valuation users $i\in \caln_h$, if any, they will be allocated $b_i$ in the offline optimal decisions. Also, with $g_i(0)=0$, $g_i(y) = \int_0^y g_i'(s)ds$. Thus, equality $(i)$ holds. For inequality $(ii)$, it can be observed that the marginal utility of any user in $\caln_l$ and $\caln_z$ (if any) is bounded by $\sum_{\ell\in\mathcal{L}_i}\phi_\ell(\omega_\ell^N)$ and the marginal utility when allocating more than $\hat{y}_i$ to user $i$ in $\caln_m$ (if any) is also bounded by $\sum_{\ell\in\mathcal{L}_i}\phi_\ell(\omega_\ell^N)$. Thus, for $i\in\caln_l\cup\caln_z$, $g_i(\Tilde{y}_i)\le\sum_{\ell\in\mathcal{L}_i}\phi_\ell(\omega_\ell^N)\Tilde{y}_i$; for $i\in\caln_m$, $g_i(\Tilde{y}_i)-g_i(\hat{y}_i)\le \sum_{\ell\in\mathcal{L}_i}\phi_\ell(\omega_\ell^N)(\Tilde{y}_i-\hat{y}_i)$. Inequality $(ii)$ then holds. The correctness of inequality $(iii)$ is based on the following observations:
\begin{itemize}
    \item For users coming after $\tau$, if any, they will be allocated nothing in the online algorithm ($y_i=0,i>\tau$), and thus their marginal utility functions shall satisfy $g'_i(y)\le \sum_{\ell\in \mathcal{L}_i}\phi_\ell(\omega_\ell^{i-1})=\sum_{\ell\in \mathcal{L}_i}\phi_\ell(\omega_\ell^N)$.
    \item The offline optimal utility reaches the largest if the users have enough budgets to occupy all the capacity left in each link and the utility functions of those users coming after $\tau$ are the largest possible $g_i'(y)=\sum_{\ell\in\mathcal{L}_i}\phi_\ell(\omega_\ell^N)-\epsilon$, where $\epsilon$ is small.
\end{itemize}

The total utility of the online algorithm under the same request sequence $\mathcal{I}$ is
\begin{equation}
\label{eq: online_expression}
\begin{split}
\text{ALG}(\mathcal{I})&=\sum_{i=1}^N g_i(y_i)= \sum_{i=1}^\tau g_i(y_i)\\
    &= \sum_{i=1}^\tau \big[g_i(y_i)-\sum_{\ell\in\mathcal{L}_i}\int_{\omega_\ell^{i-1}}^{\omega_\ell^{i}}\phi_\ell(s)ds\big] \\
    &\quad\quad+\sum_{i=1}^{\tau}\sum_{\ell\in\mathcal{L}_i}\int_{\omega_\ell^{i-1}}^{\omega_\ell^{i}}\phi_\ell(s)ds.
\end{split}
\end{equation}

Let $\Delta_i=g_i(y_i)-\sum_{\ell\in\mathcal{L}_i}\int_{\omega_\ell^{i-1}}^{\omega_\ell^i}\phi_\ell(s)ds$. Thus, $\Delta_i$ is the optimal objective value for (\ref{decision-rule}) and $\Delta_i\ge 0$. Define $\hat{\Delta}_i$ as follows: For $i\in\caln_h$, $\hat{\Delta}_i=g_i(b_i)-\sum_{\ell\in\mathcal{L}_i}\phi_\ell(\omega_\ell^N)b_i$; for $i\in\caln_m$, $\hat{\Delta}_i=g_i(\hat{y}_i)-\sum_{\ell\in\mathcal{L}_i}\phi_\ell(\omega_\ell^N)\hat{y}_i$; for $i\in\caln_l\cup\caln_z$, $\hat{\Delta}_i=0$. 
% Thus, from (\ref{eq: opt_expression}), we can have the following inequality:
% \begin{align}
% \label{eq: opt_exp2}
% \text{OPT}(\mathcal{I}) \le \sum_{i=1}^\tau \hat{\Delta}_i + \sum_{\ell\in\mathcal{L}}\phi_\ell(\omega_\ell^N).
% \end{align}
Combining (\ref{eq: opt_expression}) and (\ref{eq: online_expression}), we have
\begin{equation}
\begin{split}
\frac{\text{OPT}(\mathcal{I})}{\text{ALG}(\mathcal{I})}
    &\le \frac{\sum_{i=1}^\tau \hat{\Delta}_i + \sum_{\ell\in\mathcal{L}}\phi_\ell(\omega_\ell^N)}{\sum_{i=1}^\tau \Delta_i+\sum_{i=1}^{\tau}\sum_{\ell\in\mathcal{L}_i}\int_{\omega_\ell^{i-1}}^{\omega_\ell^N}\phi_\ell(s)ds}\\
    &\overset{(iv)}{\le} \frac{\sum_{\ell\in\mathcal{L}}\phi_\ell(\omega_\ell^N)}{\sum_{i=1}^{\tau}\sum_{\ell\in\mathcal{L}_i}\int_{\omega_\ell^{i-1}}^{\omega_\ell^N}\phi_\ell(s)ds}\\
    &= \frac{\sum_{\ell\in\mathcal{L}}\phi_\ell(\omega_\ell^N)}{\sum_{\ell\in\mathcal{L}}\int_{0}^{\omega_\ell^N}\phi_\ell(s)ds}. \label{eq: ratio}
\end{split}
\end{equation}

To show ($iv$) holds, let $h_i(\lambda)$ be the conjugate function of $g_i(y)$, i.e., $h_i(\lambda)=\max_{0\le y\le b_i}(g_i(y)-\lambda y)$. It is easy to verify that $h_i(\lambda)$ is non-increasing in $\lambda$. Notice that when $y_i>0$, i.e., $i\in \caln_h\cup\caln_m\cup\caln_l$,
\begin{align*}
   \hat{\Delta}_i &\overset{(v)}{=} h_i(\sum_{\ell\in\mathcal{L}_i}\phi_\ell(\omega_\ell^N))\overset{(vi)}{\le} h_i(\sum_{\ell\in\mathcal{L}_i}\phi_\ell(\omega_\ell^i))\\ &\overset{(vii)}{=} \max_{0\le y\le b_i}(g_i(y)-\sum_{\ell\in\mathcal{L}_i}\phi_\ell(\omega_\ell^i) y)\\ &\overset{(viii)}{\le} \max_{0\le y\le b_i} (g_i(y) - \sum_{\ell\in \mathcal{L}_i}\int_{\omega_\ell^{i-1}}^{\omega_\ell^{i-1}+y} \phi_\ell(s)ds) \overset{(ix)}{=} \Delta_i 
\end{align*}
% $\hat{\Delta}_i \overset{(j)}{=} h_i(\sum_{\ell\in\mathcal{L}_i}\phi_\ell(\omega_\ell^N))\overset{(k)}{\le} h_i(\sum_{\ell\in\mathcal{L}_i}\phi_\ell(\omega_\ell^i)) \overset{(l)}{=} \max_{0\le y\le b_i}(g_i(y)-\sum_{\ell\in\mathcal{L}_i}\phi_\ell(\omega_\ell^i) y) \overset{(m)}{\le} \max_{0\le y\le b_i} (g_i(y) - \sum_{\ell\in \mathcal{L}_i}\int_{\omega_\ell^{i-1}}^{\omega_\ell^{i-1}+y} \phi_\ell(s)ds) \overset{(n)}{=} \Delta_i$, 
where $(v)$ can be verified based on definitions of $\hat{\Delta}_i$ and $h_i(\cdot)$; $(vii)$ and $(ix)$ are based on definitions of $h_i(\cdot)$ and $\Delta_i$; $(vi)$ and $(viii)$ hold because $\phi_\ell(\omega)$ is increasing in $\omega$ and $h_i(\lambda)$ is non-increasing in $\lambda$. When $i\in\caln_z$, $\Delta_i=\hat{\Delta}_i=0$. Thus, for any $i\le\tau$, $\hat{\Delta}_i\le \Delta_i$, inequality ($iv$) holds. 

Denote the set of links where $\omega_\ell^N<\beta$ by $\mathcal{L}^1$ and the links where $\beta \le \omega_\ell^N < 1$ by $\mathcal{L}^2$, we can express (\ref{eq: ratio}) as the following:
\begin{align}
\label{eq: rewrite-ratio}
    \frac{\text{OPT}(\mathcal{I})}{\text{ALG}(\mathcal{I})} 
    &\le \frac{\sum_{\ell\in\mathcal{L}^1}m+\sum_{\ell\in\mathcal{L}^2}\phi_\ell(\omega_\ell^N)}{\sum_{\ell\in\mathcal{L}^1}m\omega_\ell^N+\sum_{\ell\in\mathcal{L}^2}\int_{0}^{\omega_\ell^N}\phi_\ell(s)ds}.
\end{align}
By the sufficient conditions (\ref{sufficient-condition-1}) and (\ref{sufficient-condition-2}), when $\omega_\ell^N\in [\beta,1]$, we have
\begin{align}\label{eq: sufficient_integral}
    \int_{0}^{\omega_\ell^N}\phi_\ell(s)ds \ge \frac{1}{\alpha}\phi_\ell(\omega_\ell^N).
\end{align}
Thus, (\ref{eq: rewrite-ratio}) can be further upper bounded as follows:
\begin{align}
    \frac{\text{OPT}(\mathcal{I})}{\text{ALG}(\mathcal{I})} 
    &\le \frac{\sum_{\ell\in\mathcal{L}^1}m+\sum_{\ell\in\mathcal{L}^2}\phi_\ell(\omega_\ell^N)}{\sum_{\ell\in\mathcal{L}^1}m\omega_\ell^N+\sum_{\ell\in\mathcal{L}^2}\frac{1}{\alpha}\phi_\ell(\omega_\ell^N)}.
    \label{eq: last-step}
\end{align}
From (\ref{eq: last-step}), the upper bound depends on $\omega_\ell^N$, and we have
\begin{align*}
    \frac{\text{OPT}(\mathcal{I})}{\text{ALG}(\mathcal{I})}
    &\le \frac{\sum_{\ell\in\mathcal{L}^1}m+\sum_{\ell\in\mathcal{L}^2}\phi_\ell(\omega_\ell^N)}{\sum_{\ell\in\mathcal{L}^2}\frac{1}{\alpha}\phi_\ell(\omega_\ell^N)} \\
    &\le \alpha\big(\frac{\lvert \mathcal{L}^1 \rvert}{\lvert \mathcal{L}^2 \rvert}+1\big)\\
    &\le \lvert\mathcal{L}\rvert\alpha = L\alpha.
\end{align*}
From the above inequalities, we observe that, in the worst case, $\omega_\ell^N=0$ for $\ell\in\mathcal{L}^1$ or $\lvert\mathcal{L}^1\rvert=0$, and there is only one link in $\mathcal{L}^2$ with $\phi_\ell(\omega_\ell^N)=m$. In summary, the worst case of arrivals in Case 2 can just contain two arrivals. The first user will only request one single link $\ell$ with utility satisfying $g'_1(y)=m$, followed by the second user requesting all links with utility satisfying $g'_2(y)= m|\mathcal{L}|$. Both of their budgets are greater than $\beta$. The online algorithm will only accept the first user and allocate $\beta$ of link $\ell$ to him, producing a total utility of $\beta m$. The offline optimal will accept the second user with the total utility being $m|\mathcal{L}|$. 

In Case 3, there exist links with the capacity limit reached. $\mathcal{L}^1$ and $\mathcal{L}^2$ are defined in the same way as that in Case 2 and the links with $\omega_\ell^N=1$ are denoted by $\mathcal{L}^3$. We do not need to group users by their utility functions as in Case 2, because for any user $i$, $g'_i(y)\le \lvert\mathcal{L}_i\rvert M$, and each link carries a capacity limit of 1, we can directly bound the offline optimal as follows:
\begin{align*}
    \text{OPT}(\mathcal{I}) \le \lvert\mathcal{L}\rvert M.
\end{align*}
The utility of the online algorithm is lower bounded as follows:
\begin{align*}
    \text{ALG}(\mathcal{I}) &\ge \sum_{\ell\in\mathcal{L}^1}m\omega_\ell^N+\sum_{\ell\in\mathcal{L}^2 \cup \mathcal{L}^3}\int_{0}^{\omega_\ell^N}\phi_\ell(s)ds \\
    &\overset{(x)}{\ge} \sum_{\ell\in\mathcal{L}^1}m\omega_\ell^N + \frac{1}{\alpha}\sum_{\ell\in\mathcal{L}^2}\phi_\ell(\omega_\ell^N) + \frac{1}{\alpha}\sum_{\ell\in\mathcal{L}^3}M,
\end{align*}
where $(x)$ follows from equation~\eqref{eq: sufficient_integral}. Then we have
\begin{align*}
    \frac{\text{OPT}(\mathcal{I})}{\text{ALG}(\mathcal{I})} &\le \frac{\lvert\mathcal{L}\rvert M}{\sum_{\ell\in\mathcal{L}^1}m\omega_\ell^N + \frac{1}{\alpha}\sum_{\ell\in\mathcal{L}^2}\phi_\ell(\omega_\ell^N) + \frac{1}{\alpha}\sum_{\ell\in\mathcal{L}^3}M}\\
    &\overset{(xi)}{\le} \alpha\lvert\mathcal{L}\rvert=L\alpha.
\end{align*}
We observe that for inequality ($xi$) to be equal, we must have $\lvert\mathcal{L}^1\rvert=0$ and $\lvert\mathcal{L}^2\rvert=0$.

Thus, combining results in Cases 1-3, the online algorithm we proposed is $L\alpha$-competitive, linear in the number of links.
\end{proof}

In the sequel, we show that our analysis above is tight by explicitly presenting the worst-case instances for the algorithm to reach the competitive ratio claimed. We have presented a special worst-case instance with two arrivals for Case 2 in the proof of Theorem \ref{thm: suff_conditions}. A more general property of the worst-case instances for $\text{OA}_\phi$ is shown in the following lemma.

\begin{lemma}
Any worst-case instance for $\text{OA}_\phi$ must possess the property: for each arrival $i<\tau$, $g_i(y_i)= \sum_{\ell\in\mathcal{L}_i}\int_{\omega_\ell^{i-1}}^{\omega_\ell^N}\phi_\ell(s)ds+\epsilon$.
\end{lemma}

\noindent
\begin{proof}
Notice that the inequality ($iv$) in equation~\eqref{eq: ratio} becomes equal only if the arrival instance $\mathcal{I}$ owns the property above. We prove the lemma by constructing a general arrival sequence $\mathcal{I}$ such that $\frac{\text{OPT}(\mathcal{I})}{\text{ALG}(\mathcal{I})}=L\alpha$. 

The arrival instance is constructed as follows: there comes a group of $t_1$ arrivals requesting a specific link $\ell$, whose utility functions satisfy $$\forall y\in[0,b_i],g'_i(y)/\lvert\mathcal{L}_i\rvert=g'_i(y)=m,$$ and the budgets satisfy $b_i=\frac{1}{\alpha t_1}$, $i=1,\dots,t_1$; followed by a second group of $t_2$ arrivals requesting the same link $\ell$, whose utility functions satisfy $$\exists y_i \text{ close to }0,g'_i(y_i)=g'_{i-1}(y_i)+\frac{\epsilon}{t_2},$$ and the budgets $b_i=1$, $i=t_1+1,\dots,t_1+t_2$; at last, an arrival comes and requests all the links, whose utility function satisfies $$g'_{t_1+t_2+1}(y)=mL+\frac{\epsilon}{2},$$ and $b_{t_1+t_2+1}=1$. 

For the first group of arrivals, the online algorithm will 
allocate rate $y_i=b_i$ and the utilization of link $\ell$ reaches $\frac{1}{\alpha}$ before the second group. In the sequel, for arrival $i=t_1+1$, the online algorithm will allocate rate $$y_i=\phi_\ell^{-1}(m+\frac{(i-t_1)\epsilon}{t_2})-\frac{1}{\alpha}.$$ 
For arrivals $i=t_1+2, \dots, t_1+t_2$, the online algorithm will allocate rate \begin{align*}
    y_i&=\phi_\ell^{-1}(m+\frac{(i-t_1)\epsilon}{t_2})-\phi_\ell^{-1}(m+\frac{(i-t_1-1)\epsilon}{t_2})\\
    &\approx \phi'^{-1}_\ell(m+\frac{(i-t_1)\epsilon}{t_2})\frac{\epsilon}{t_2}.
\end{align*} 
For arrival $i=t_1+t_2+1$, namely, the last arrival comes, we have $$\sum_{\ell\in\mathcal{L}}\phi_\ell(\omega_\ell^{t_1+t_2})=mL+\epsilon > g'_{t_1+t_2+1}(0),$$ and thus the solution to \eqref{decision-rule} is $y_{t_1+t_2+1}=0$, i.e., the online algorithm will reject the last arrival. 

However, as long as the number of links $L>1$, in terms of the capacity allocation of the frequently requested link $\ell$, rates assigned to the last arrival obviously yields more utility than those assigned to the first two groups since the last arrival requests all the links. Thus, the offline optimal will reject the first two groups and allocate everything to the last arrival. This being the case, we have
\begin{align*}
    \frac{\text{OPT}(\mathcal{I})}{\text{ALG}(\mathcal{I})}&=\frac{mL+\frac{\epsilon}{2}}{\sum_{i=1}^{t_1}g_i(\frac{1}{\alpha t_1})+\sum_{i=t_1+1}^{t_1+t_2}g_i(y_i)} \\
    &= \frac{mL+\frac{\epsilon}{2}}{\frac{m}{\alpha}+\sum_{i=t_1+1}^{t_1+t_2}\int_0^{y_i}g'_i(y)dy}.
\end{align*}
Thus, let $t_1,t_2\rightarrow\infty,\epsilon\rightarrow 0$, we have
\begin{align*}
    \frac{\text{OPT}(\mathcal{I})}{\text{ALG}(\mathcal{I})}\rightarrow\frac{mL}{\frac{m}{\alpha}+m\cdot\phi_\ell'^{-1}(m)\cdot 0}=L\alpha,
\end{align*}
which shows that our analysis is tight.
\end{proof}

\section{Simulation Results} \label{sec:simulation}
% Simulations

\subsection{Simulation Settings}
We use the network trace of the Abilene network collected from December 8, 2003 to December 28, 2003 to demonstrate the performance of our algorithm~\cite{dataset}. There are 11 nodes, 41 links and 110 source-destination pairs in the Abilene network. The network trace contains the traffic matrix and the routing matrix. The traffic matrix given in the data set is a real-valued matrix with dimension $2016\times 110$, where the component on the $i$th column and the $j$th row represents the traffic volume of the $i$th source-destination pair measured in the $j$th 5-min slot. The routing matrix is a binary matrix whose dimension is $110\times 41$, where the $i$th row vector denotes the routing path of the corresponding source-destination pair.

The arrival order is obtained based on the traffic matrix. We order the link by the traffic measured in each 5-min slot and assume that the arrivals on a specific link in a 5-min slot are generated by a Poisson process, the mean arrival rate of which is proportional to the traffic amount of that link. Note that the Poisson assumption is not necessary to our theoretical analysis but only adopted for generating the synthetic arrival data. We obtain the requested link set of each arrival based on the routing matrix. Utilities and budgets are not included in the data set. We assume that the utility functions of arrivals are in the form of $g_i(y)=a_i|\mathcal{L}_i|\log (1+y)$, where $\mathcal{L}_i$ is the link set requested by the $i$th arrival. We consider two cases for $a_i$: the average case and the worst case. In the average case, $a_i$s are randomly generated from a truncated Gaussian distribution, $a_i \sim \text{TruncGaussian}(\mu,\sigma,LB,UB)$, where $\mu,\sigma,LB,UB$ are the mean, variance, lower bound and upper bound, respectively. The mean and variance are fixed for each instance and the bounds are properly set so that $\frac{g'_i(y)}{|\mathcal{L}_i|}\in [m,M]$. In the worst case, $a_i \sim \text{TruncGaussian}(\mu_i,\sigma_i,LB,UB)$, with means $\mu_i$ slowly increasing or decreasing with $i$. We assume that the budgets of arrivals are uniformly distributed $b_i \sim U(0,1)$.

Before this work, there is no algorithm designed for our ONUM problem. Thus, we compare our proposed algorithm with two heuristics, which are greedy and reservation-based algorithm. The descriptions of the heuristics are as follows:

% \begin{itemize}
\textbf{Greedy.} $y_i=b_i$ as long as the capacity permits, otherwise $y_i = \min_{\ell\in\mathcal{L}_i}(1-\omega^{i-1}_\ell)$, where $\omega^i_\ell$ is the utilization level of link $\ell$ after the $i$th allocation $y_i$ is finished. The implementation of this algorithm will be referred to as \textbf{greedy} hereinafter.

\textbf{Reservation-based algorithm.} Reserve $p$ of the capacity for arrivals with $\frac{g'_i(y)}{|\mathcal{L}_i|}\ge q M$, where $p,q\in [0,1]$ are parameters to be determined. It is plain to see that it is beneficial for high-utility arrivals if $\alpha,\beta$ are closer to 1 and beneficial for low-utility arrivals otherwise. This class of algorithms are widely used in network control problems\cite{key_1990}. The implementations of this algorithm will be referred to as \textbf{res} or \textbf{reservation} hereinafter.
% \end{itemize}

\subsection{Results of Plain Implementations}\label{subsec:plain}
% Figure 1: utility-mean(a), utility-variance(b). Notation: performance in the average case 
% Figure 2: utility-hl(a), utility-lh(b). Notation: performance in the worst case

\begin{figure}
\centering
\begin{subfigure}{\columnwidth}
  \centering
  \includegraphics[width=\linewidth]{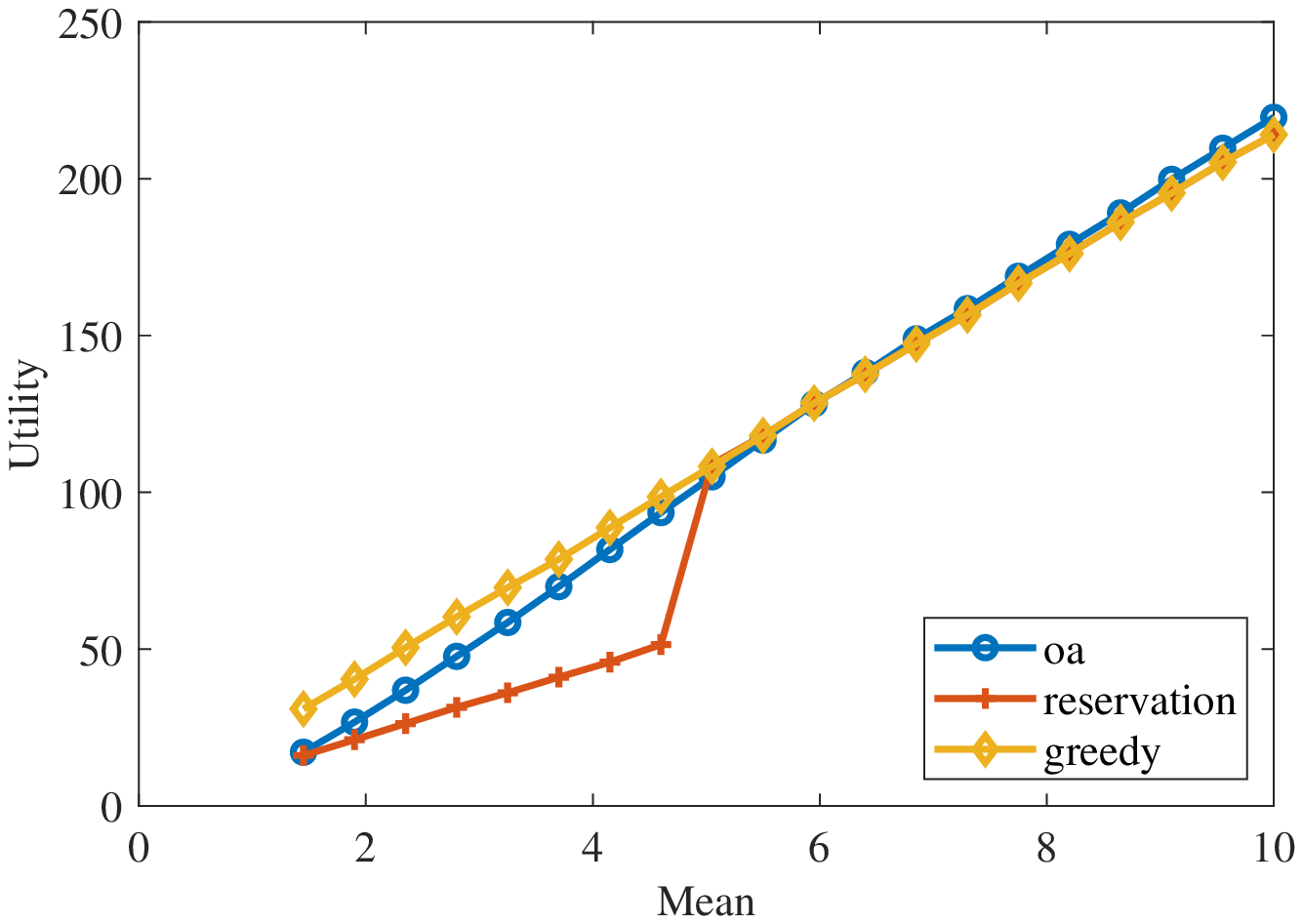}
  \caption{Utility vs. Mean (Var=0.1)}
  \label{fig:avr-uvsm-v01}
\end{subfigure}
\begin{subfigure}{\columnwidth}
  \centering
  \includegraphics[width=\linewidth]{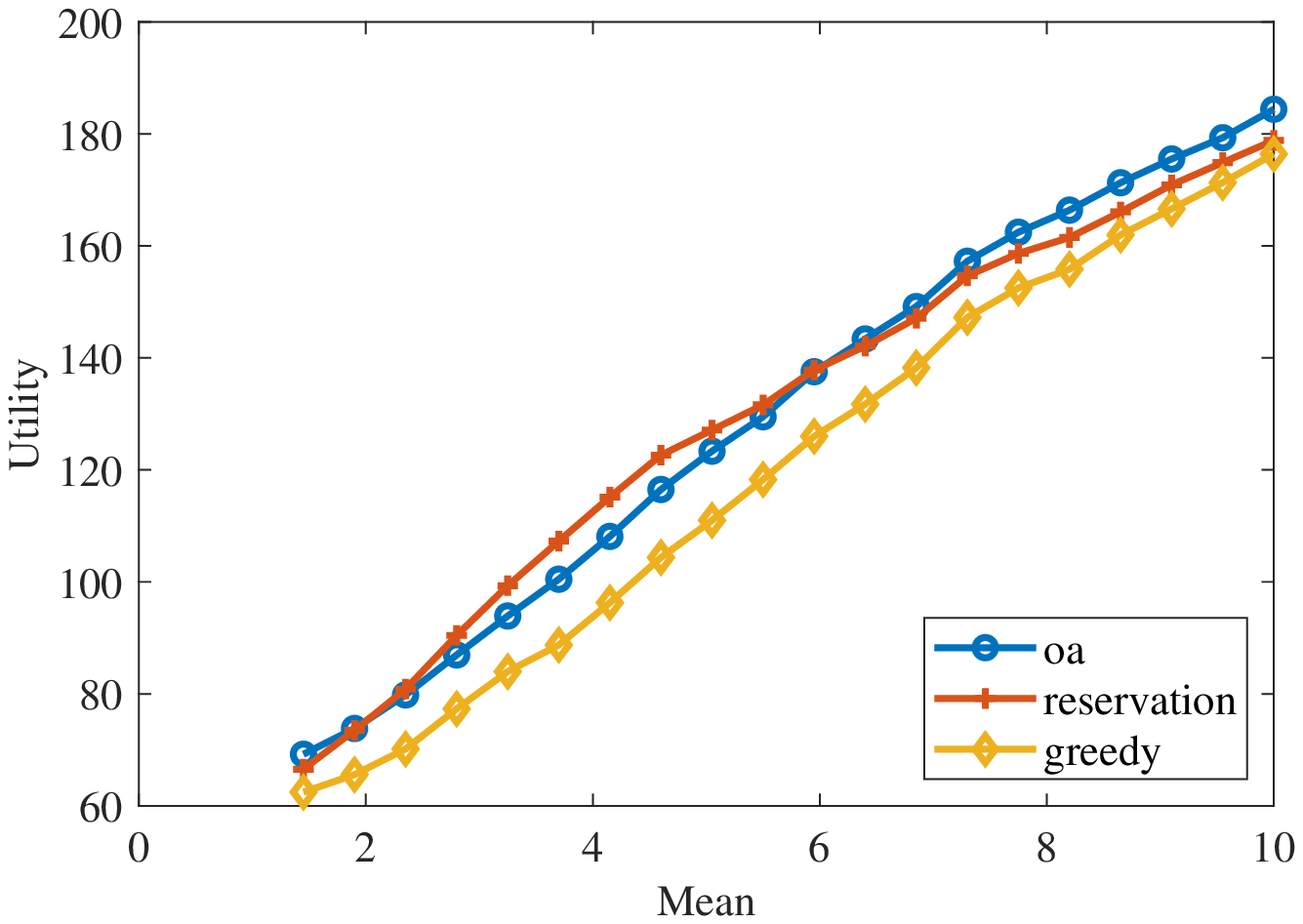}
  \caption{Utility vs. Mean (Var=3)}
  \label{fig:avr-uvsm-v3}
\end{subfigure}
\caption{Performance of different algorithms with plain implementation in the average case. The utilities of arrivals are generated by a truncated Gaussian distribution. The $y$-axis is the average utility produced by algorithms and the $x$-axis is the mean of the distribution. Two sub-figures show the performance under arrivals with low and high uncertainties.}
\label{fig:vanilla-avr-case}
\end{figure}

\begin{figure}
\centering
\begin{subfigure}{\columnwidth}
  \centering
  \includegraphics[width=\linewidth]{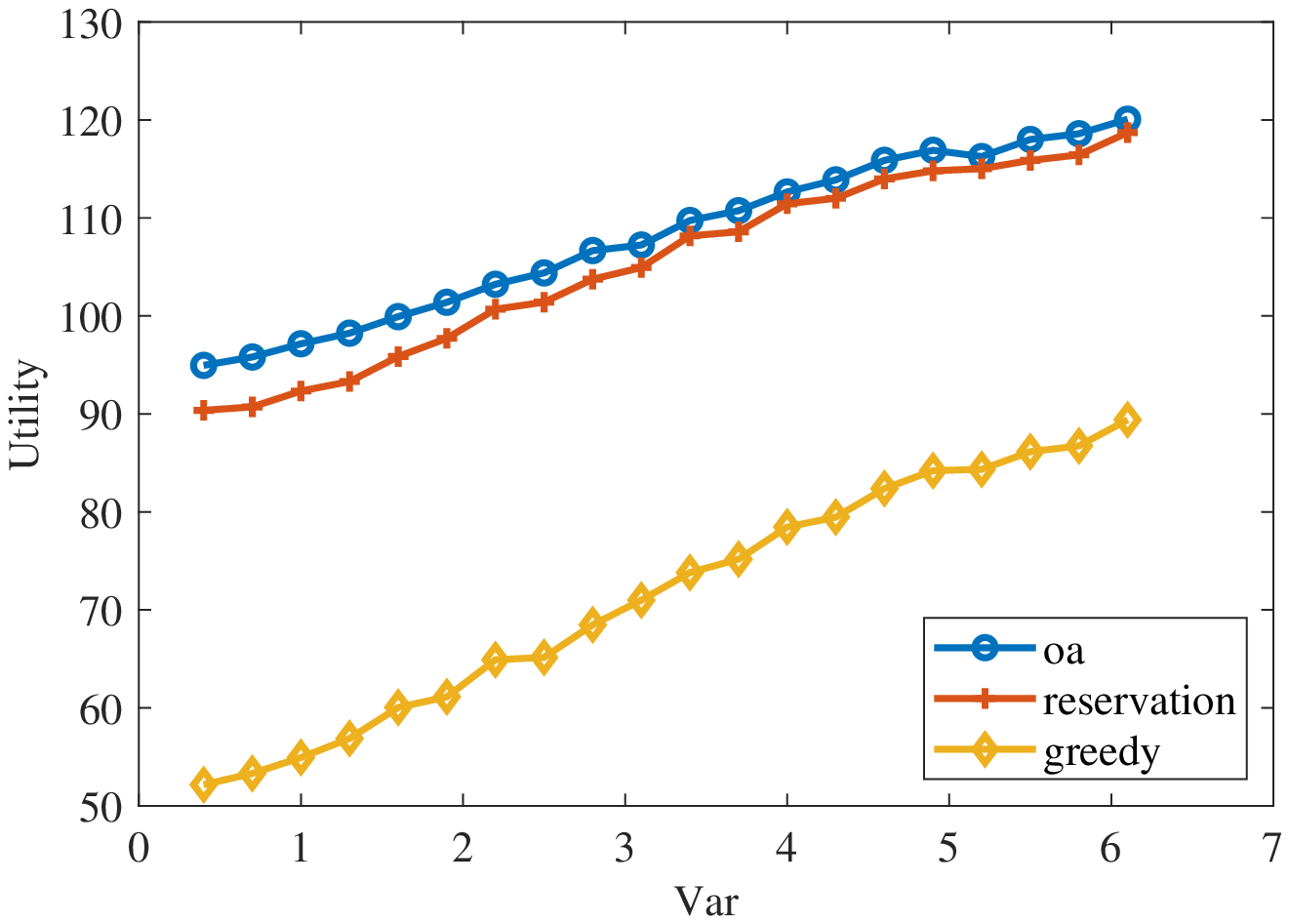}
  \caption{Utility vs. Var (increasing mean)}
  \label{fig:ext-increasing}
\end{subfigure}
\begin{subfigure}{\columnwidth}
  \centering
  \includegraphics[width=\linewidth]{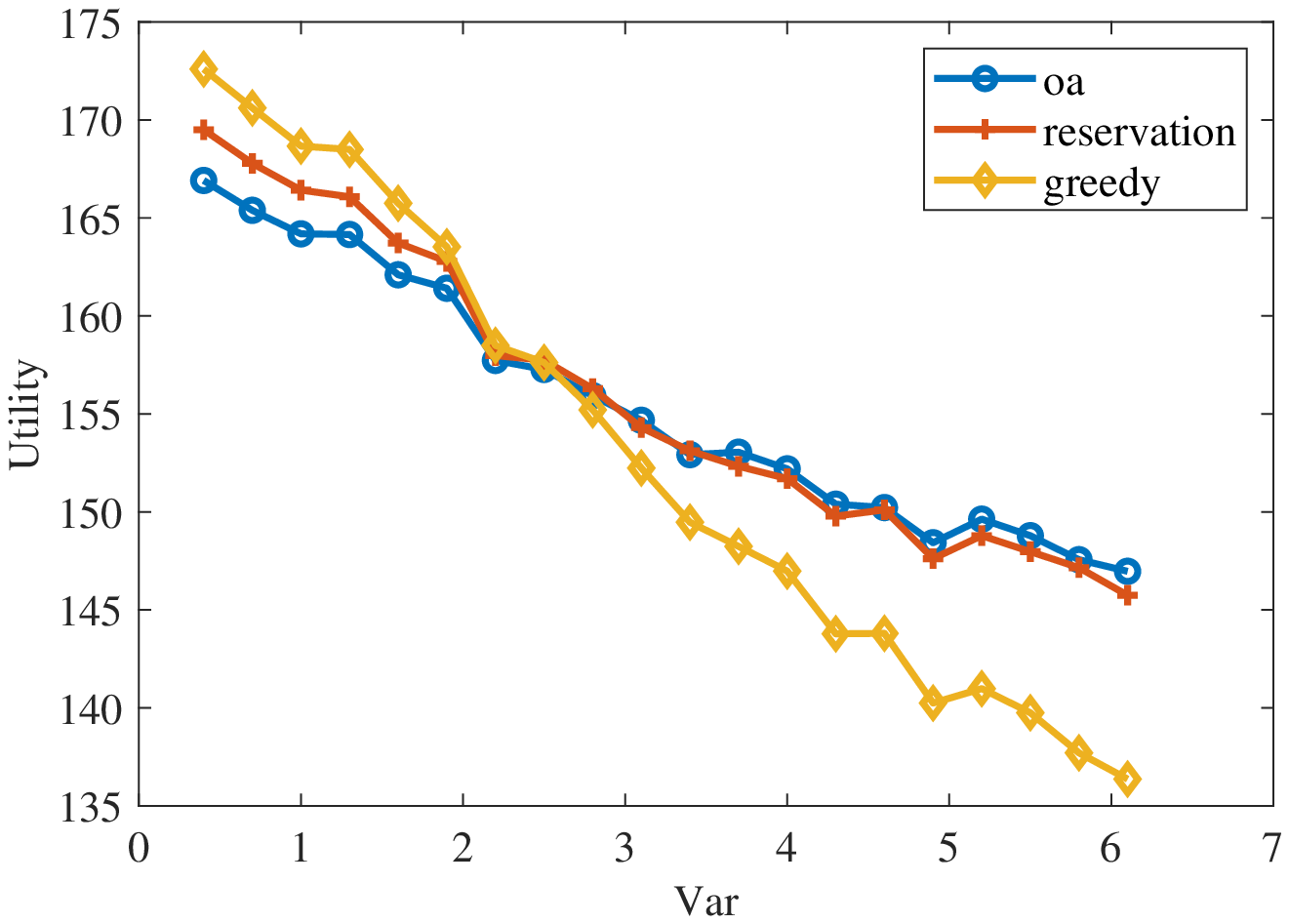}
  \caption{Utility vs. Var (decreasing mean)}
  \label{fig:ext-decreasing}
\end{subfigure}%
\caption{Performance of different algorithms with plain implementation in the extreme case. The utilities of arrivals are either increasing or decreasing on average. The $x$-axis is the uncertainty of the arrivals. The larger Var, the more uncertain the utilities are. Two sub-figures show the performance when the utilities are in general in an increasing or decreasing trend.}
\label{fig:vanilla-ext-case}
\end{figure}

Fig. \ref{fig:vanilla-avr-case} shows the total utility achieved by the aforementioned three algorithms w.r.t. the mean or variance of arrival instances. From Fig. \ref{fig:avr-uvsm-v01}, when the utilities are densely centered around a certain value, i.e., the variance is small, our algorithm performs stable with the mean while we observe a sharp twist from the reservation-based algorithm. The position of the twist in reservation-based algorithm relies badly on the parameter chosen and our algorithm avoids such instability by the virtue of the smooth price function. Fig. \ref{fig:avr-uvsm-v3} shows the case when the utilities expose higher uncertainty, our algorithm performs the best among the three when high-utility arrivals are the mainstream. Meanwhile, when the arrivals are coming with a low utility on average, i.e., the mean is small, our algorithm is not so competitive in both cases because we are too conservative to accept enough low-utility arrivals. It leaves us the space to further improve our algorithm that will be shown in Section \ref{subsec: adaptive}.  

Fig. \ref{fig:vanilla-ext-case} shows the total utility of the three algorithms w.r.t. the variance while the utilities are slowly increasing/decreasing on average with time. Our algorithm always produces a higher utility than the other two no matter of the variance when the utilities are increasing, which is the hard case for online algorithm in general, because decisions must be made more conservatively so as to reserve capacity for future users. Fig. \ref{fig:ext-increasing} validates our conjecture, where our algorithm yields the highest utility. Fig. \ref{fig:ext-decreasing} shows that, when the utilities are decreasing on average, if the utility realizations are in fact decreasing, i.e., the variance is small, it is more beneficial to be aggressive, as shown by the advantage of the greedy algorithm. Actually, all three algorithms allocate most of the capacity in the beginning; however, our algorithm is the most pessimistic because of its worst-case nature. Thus, it always waits for better deals in the future, which is impossible in this extreme case, and thus it is sub-optimal when the variance is small in Fig. \ref{fig:ext-decreasing}. In the sequel, we will show that if introducing a proper parameter-choosing mechanism, our algorithm will perform the best across all possible cases.

\subsection{Results of Adaptive Implementations}\label{subsec: adaptive}
\begin{figure}
\centering
\begin{subfigure}{\columnwidth}
  \centering
  \includegraphics[width=\linewidth]{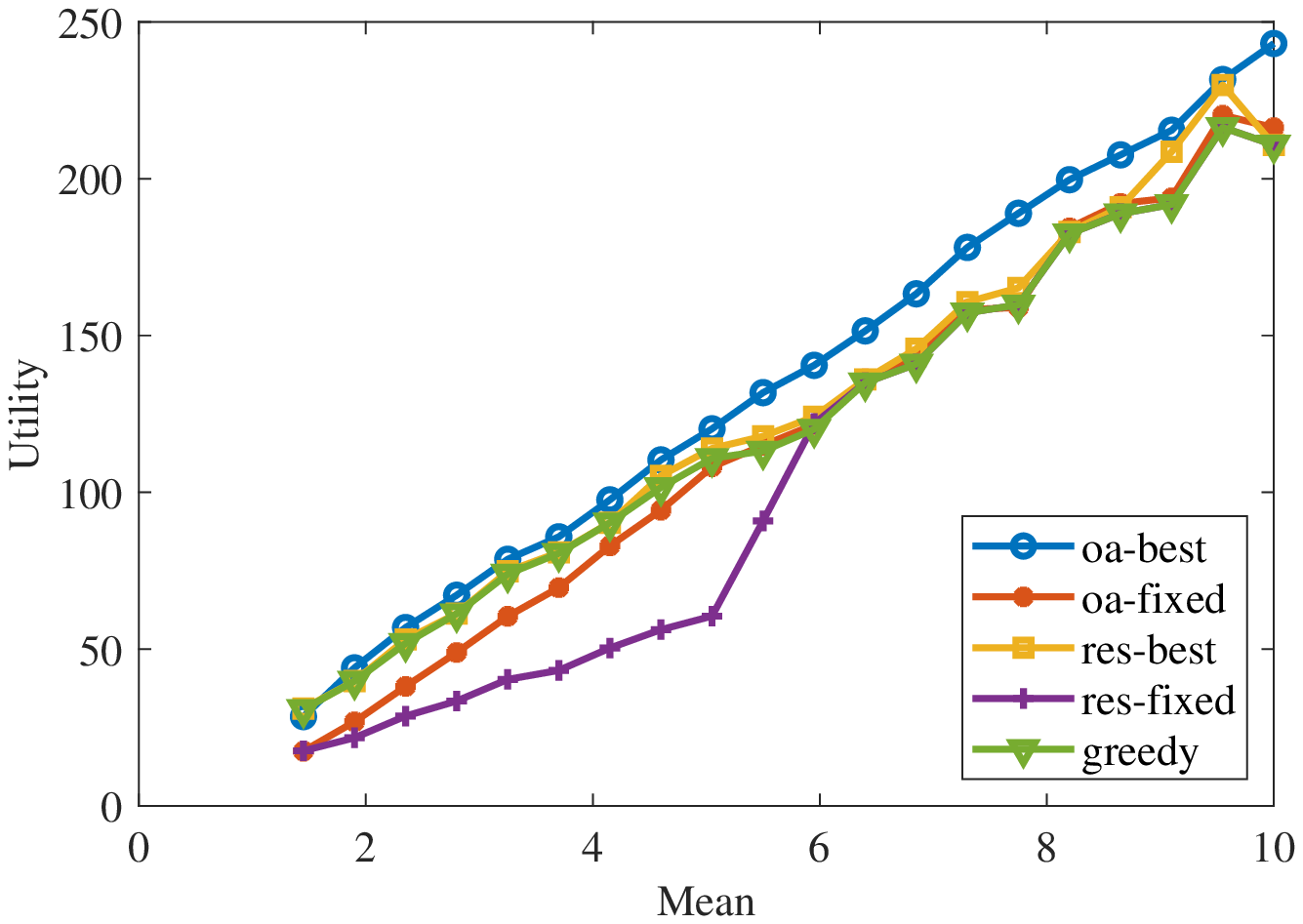}
  \caption{Utility vs. Mean (Var=0.1)}
  \label{fig:pre-avr-uvsm-v01}
\end{subfigure}
\begin{subfigure}{\columnwidth}
  \centering
  \includegraphics[width=\linewidth]{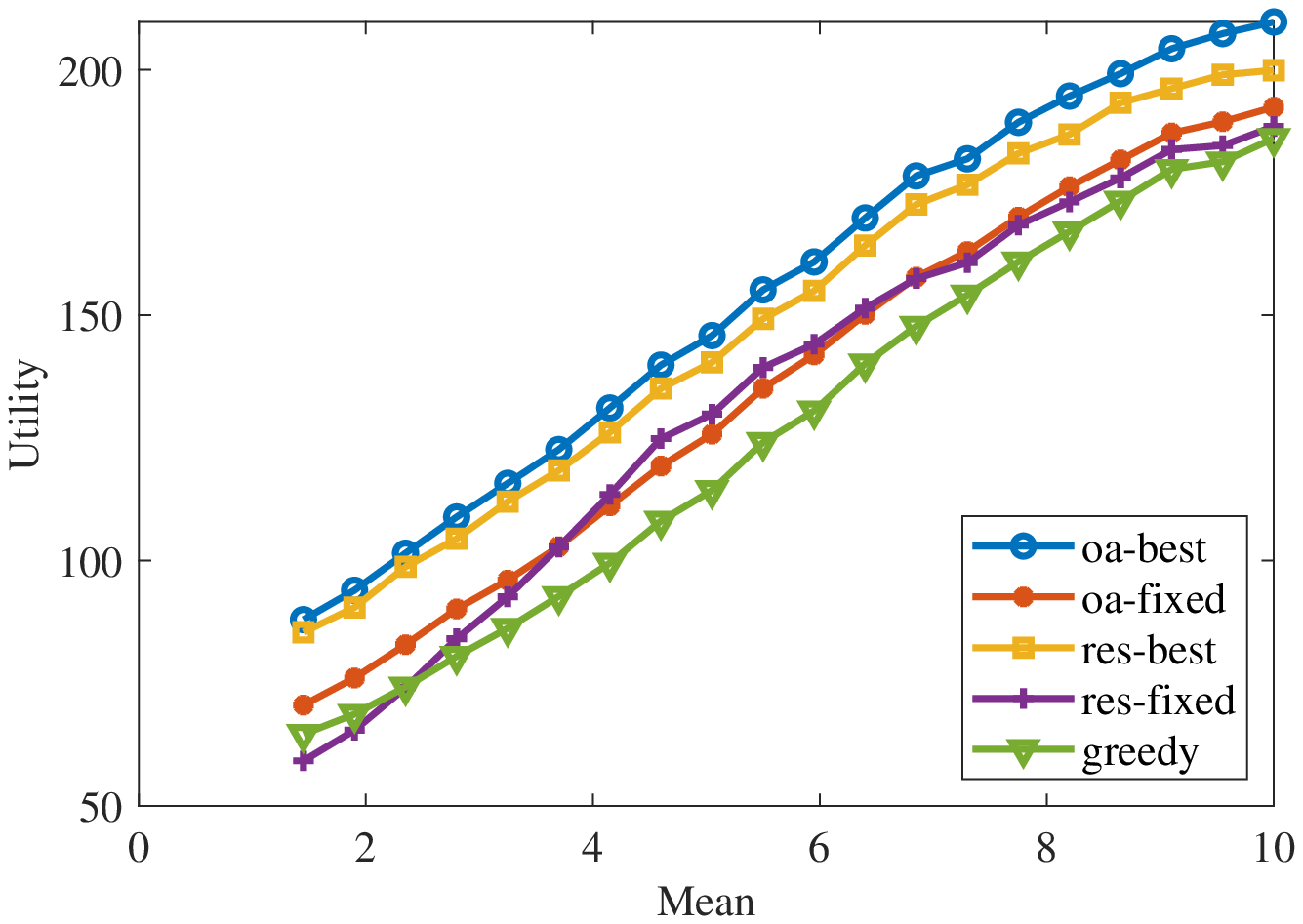}
  \caption{Utility vs. Mean (Var=3)}
  \label{fig:pre-avr-uvsm-v3}
\end{subfigure}
\caption{Performance improvement of algorithms in the average case when the parameters are chosen offline. The arrivals are generated in a similar way to Fig. \ref{fig:vanilla-avr-case}. \textbf{Alg-best} denotes the performance with the best set of parameters and \textbf{Alg-fixed} denotes the performance with the same set of parameters in Fig. \ref{fig:vanilla-avr-case}.}
\label{fig:pre-avr-case}
\end{figure}

\begin{figure}
\centering
\begin{subfigure}{\columnwidth}
  \centering
  \includegraphics[width=\linewidth]{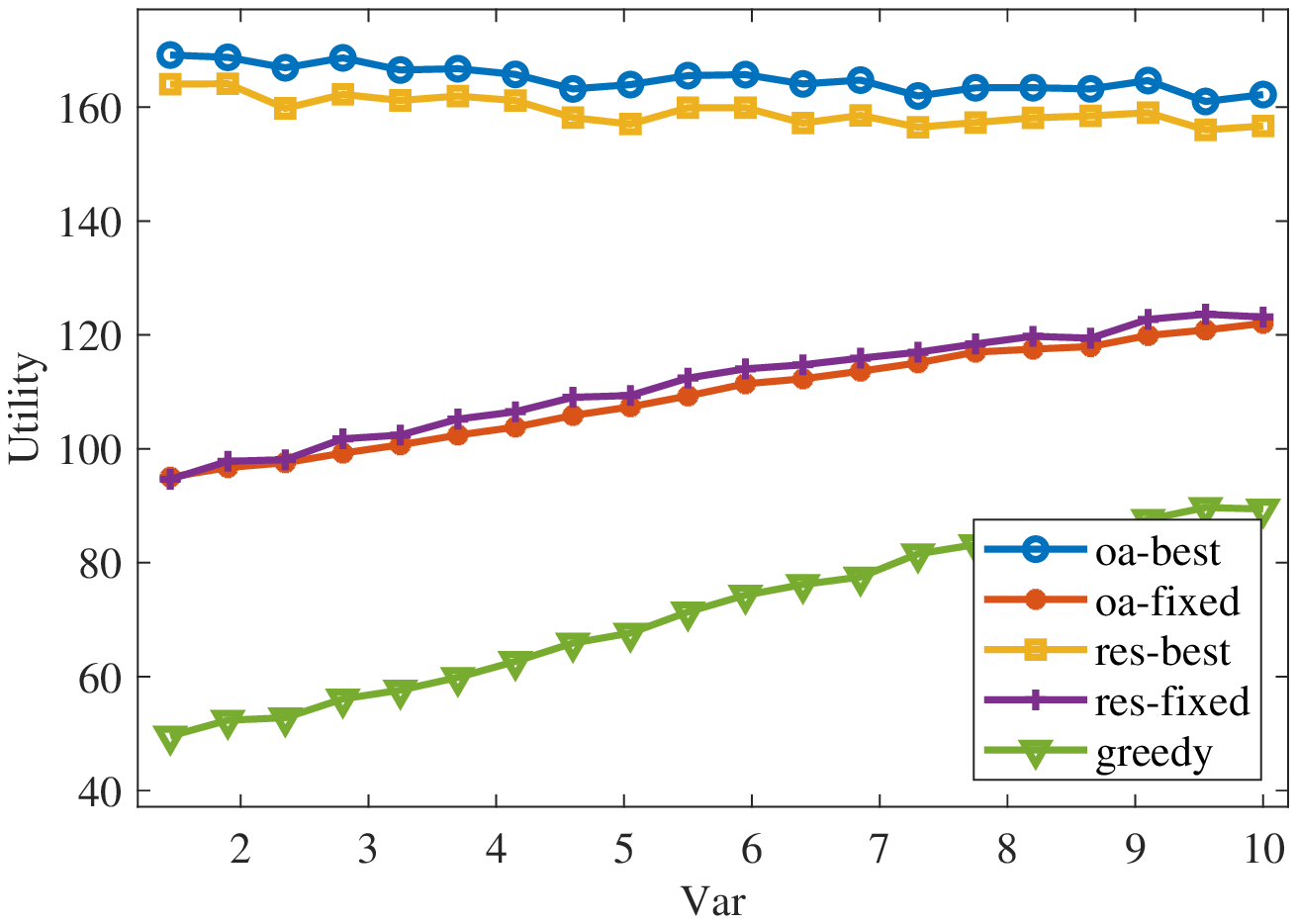}
  \caption{Utility vs. Var (increasing mean)}
  \label{fig:pre-ext-increasing}
\end{subfigure}
\begin{subfigure}{\columnwidth}
  \centering
  \includegraphics[width=\linewidth]{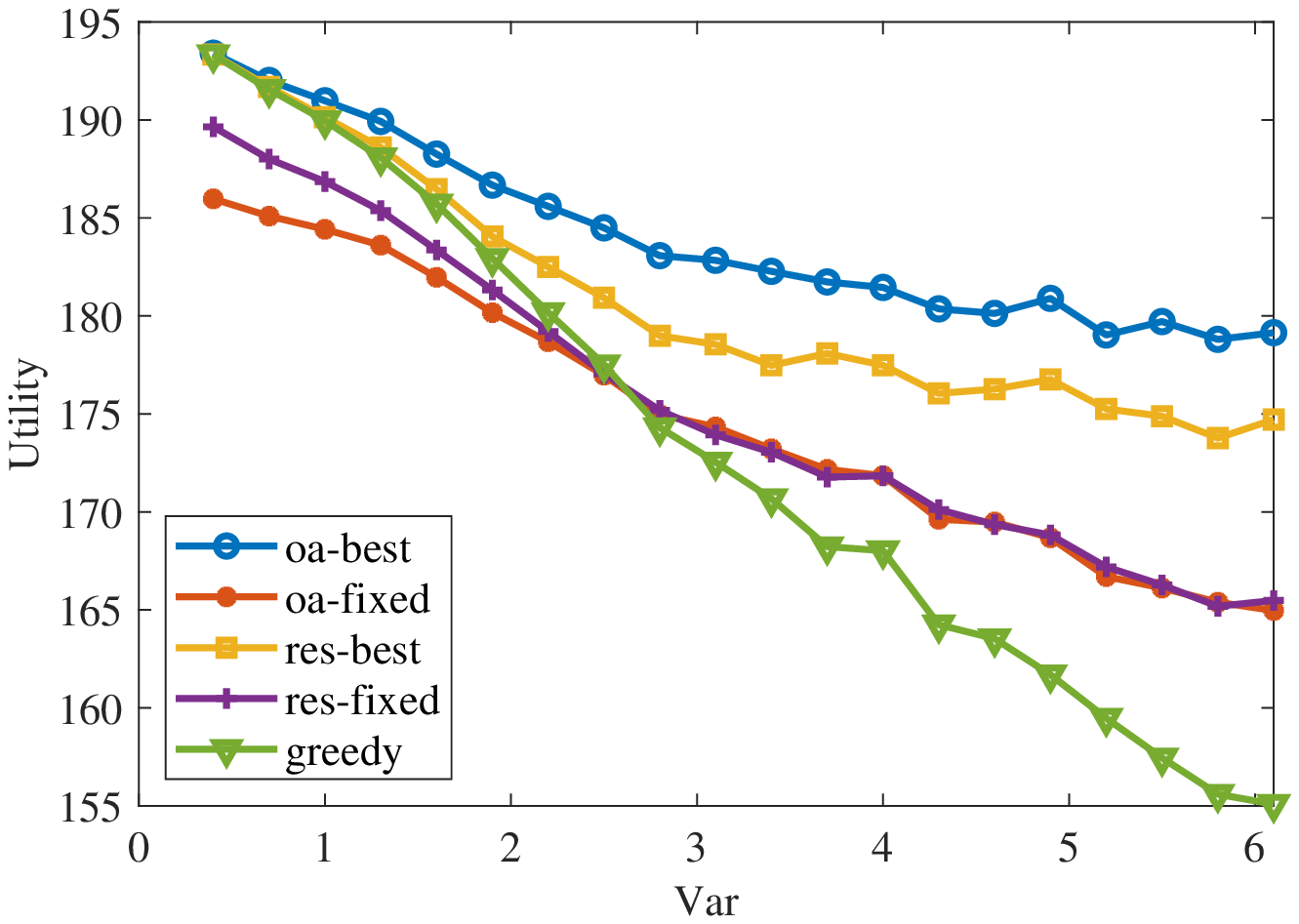}
  \caption{Utility vs. Var (decreasing mean)}
  \label{fig:pre-ext-decreasing}
\end{subfigure}%
\caption{Performance improvement of algorithms in the extreme case when the parameters are chosen offline. The arrivals are generated in a similar way to Fig. \ref{fig:vanilla-ext-case}. \textbf{Alg-best} denotes the performance with the best set of parameters and \textbf{Alg-fixed} denotes the performance with the same set of parameters in Fig. \ref{fig:vanilla-ext-case}.}
\label{fig:pre-ext-case}
\end{figure}
In Section \ref{subsec:plain}, our algorithm is at a disadvantage under certain arrival instances. Actually, this is not uncommon for algorithms designed by the worst-case analysis. Those algorithms are completely agnostic to the input patterns, and thus, if the input indeed poses a pattern such as some stochasticity, algorithms designed for taking care of the worst-case scenarios will fail pathetically. However, we are going to show that with the right parameters, our algorithm is able to be both theoretically and practically competitive. To accomplish that, we need to understand what information our algorithm fails to capture in the instances. We observe that the common feature of the instances under which our algorithm is disadvantageous is that either $m$ or $M$ serves as a loose bound of $\frac{g'_i(y)}{|\mathcal{L}_i|}$; in other words, the utility bounds that our algorithm uses in Algorithm 1 are not tight enough. A conjecture is that if such pattern is incorporated into the algorithm, our algorithm will be able to overcome its intrinsic worst-case nature and yield competitive performance even under average cases. Therefore, we first confirm our conjecture by doing a simple experiment. We replace the original $(m,M)$ by $(\Tilde{m},\Tilde{M})$ to indicate that they are variables to be chosen. A group of arrival instances are generated by a certain distribution. We run parallel implementations of our algorithm with different $(\Tilde{m},\Tilde{M})$ under those arrival instances generated, and apply our algorithm with the $(\Tilde{m},\Tilde{M})$ pair producing the most utility on average to instances newly generated by the same distribution. If our algorithm with the parameters learned offline is able to perform better over other algorithms, then our conjecture is proved. The utilities in Fig. \ref{fig:pre-avr-case} are generated from the same distribution used in Fig. \ref{fig:vanilla-avr-case}. The possible parameter space is a convex region $\{(\Tilde{m},\Tilde{M})|m\le \Tilde{m} \le \Tilde{M} \le M\}$. We discretize the space into grids with grid size $\frac{M-m}{20}\times\frac{M-m}{20}$. Although the discretization can lead to suboptimality in the parameters, a smaller search space enables a faster exhaustive search. Since we are expecting to observe the performance gain from a better parameter choice instead of optimizing the parameters, the upside dominates the downside. As shown in Fig. \ref{fig:pre-avr-case} and Fig. \ref{fig:pre-ext-case}, if parameters $(\Tilde{m},\Tilde{M})$ are chosen offline by exhaustive search, our algorithm yields the most utility among all candidate algorithms.

% \begin{figure}
%   \centering
%   \includegraphics[width=0.5\linewidth]{./pics/para-region}
%   \caption{Parameter Region}
%   \label{fig:l-u-region}
% \end{figure}%

Though the method above is effective, we may not have enough offline data to learn the parameters accurately. An alternate idea is to adaptively tune $m$ and $M$ as the algorithm runs. One way is to model the tuning process as a multi-armed bandit problem and treat different $(\Tilde{m},\Tilde{M})$ as potential actions. The expected reward of action $(\Tilde{m},\Tilde{M})$ is the utility gained by choosing $(\Tilde{m},\Tilde{M})$ as the parameters in the value function $\phi_\ell(y)$ in one episode. The algorithm then chooses arm $(\Tilde{m},\Tilde{M})$ based on the utility observed. However, for our problem, full feedback is possible if parallel computing is employed. Thus, we model the tuning process as an expert problem instead, which enjoys a faster convergence. We adopt the exponential weight algorithm proposed in \cite{auer2002nonstochastic} and show the complete algorithm in Algorithm \ref{alg:exp-weight}.

\begin{figure}
\centering
\begin{subfigure}{\columnwidth}
  \centering
  \includegraphics[width=\linewidth]{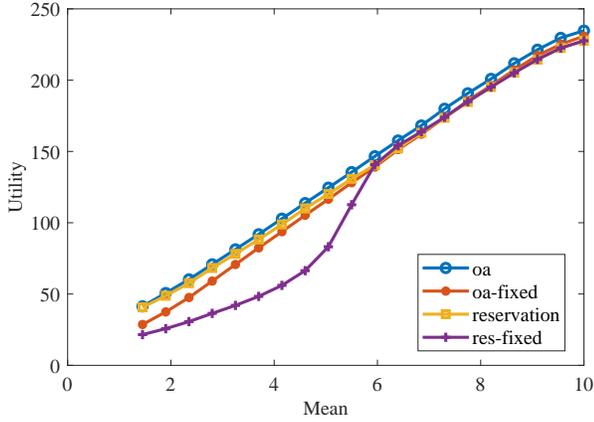}
  \caption{Utility vs. Mean (Var=0.1)}
  \label{fig:adp-avr-uvsm-v01}
\end{subfigure}
\begin{subfigure}{\columnwidth}
  \centering
  \includegraphics[width=\linewidth]{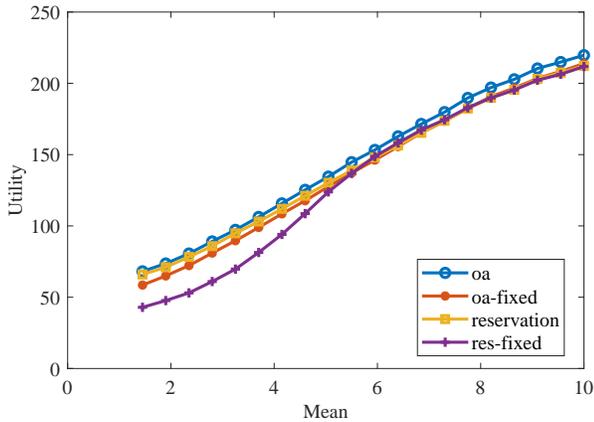}
  \caption{Utility vs. Mean (Var=3)}
  \label{fig:adp-avr-uvsm-v3}
\end{subfigure}
\caption{Performance improvement of algorithms with learned parameters in the average case. The arrivals are generated in a similar way to Fig.\ref{fig:vanilla-avr-case}. \textbf{Alg} denotes the performance with the learned parameters and \textbf{Alg-fixed} denotes the performance with the same set of parameters in Fig. \ref{fig:vanilla-avr-case}.}
\label{fig:adp-avr-case}
\end{figure}

The action space is the same as the discretized parameter space mentioned before. In Algorithm \ref{alg:exp-weight}, each episode consists of arrivals in 24 hours, which are generated by 288 rows in the traffic matrix, since each row represents the traffic amount measured in 5 min. We denote by $e_i$ the number of arrivals in the $i$th episode. We set $E=30$ and $\eta=10$ in the simulation. We show that the adaptive implementation indeed improves the performance under the previous disadvantageous instances.

\begin{figure}
\centering
\begin{subfigure}{\columnwidth}
  \centering
  \includegraphics[width=\linewidth]{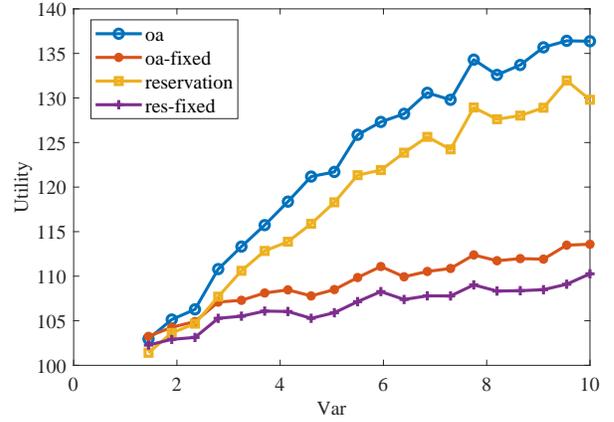}
  \caption{Utility vs. Var (increasing mean)}
  \label{fig:adp-ext-inc}
\end{subfigure}
\begin{subfigure}{\columnwidth}
  \centering
  \includegraphics[width=\linewidth]{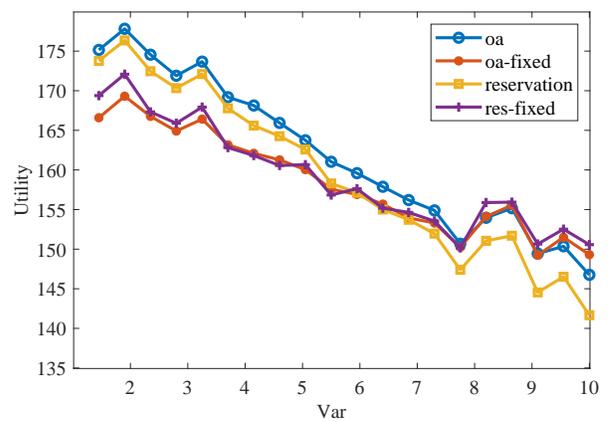}
  \caption{Utility vs. Var (decreasing mean)}
  \label{fig:adp-ext-dec}
\end{subfigure}
\caption{Performance improvement of algorithms with learned parameters in the extreme case. The arrivals are generated in a similar way to Fig.\ref{fig:vanilla-ext-case}. \textbf{Alg} denotes the performance with the learned parameters and \textbf{Alg-fixed} denotes the performance with the same set of parameters in Fig. \ref{fig:vanilla-ext-case}.}
\label{fig:adp-ext-case}
\end{figure}

\begin{algorithm}
\caption{Online Adaptive Implementation of Algorithm 1}
\begin{algorithmic}[1]\label{alg:exp-weight}
\Initialize Initial utilization $\omega_\ell^0=0$.  Weights $w_{(\Tilde{m},\Tilde{M})}=w_0,(\Tilde{m},\Tilde{M})\in\mathbb{A} $, initial probability of choosing arm $(\Tilde{m},\Tilde{M})$ is $p_{(\Tilde{m},\Tilde{M})}=w_{0}/(\sum_{(\Tilde{m},\Tilde{M})} w_{0})=1/|\mathbb{A}|$. Episode length $e_i$ of episode $i$.
\FOR{the $i$th episode}
\STATE Choose $(\Tilde{m}_i,\Tilde{M}_i)$ according to the probability distribution $\mathbf{p}$.
\STATE Initialize the link utilization levels to 0.
\FOR{the $t$th arrival ($t=(i-1)e_i+1:i e_i$)} 
\STATE Observe $A_i=\{g_i(\cdot),\mathcal{L}_i,b_i\}$.
\STATE Determine $y_i$ by solving problem (\ref{decision-rule}) with $(\Tilde{m}_i,\Tilde{M}_i)$ replacing $(m,M)$ in $\phi$ and accumulate utility $u_{(\Tilde{m},\Tilde{M})}=u_{(\Tilde{m},\Tilde{M})}+g_i(y_i)$.
\STATE Update the link utilization levels: $\omega_\ell^t = \omega_\ell^{t-1}+y_i, \ell\in\mathcal{L}_i.$
\ENDFOR
\STATE Simultaneously observe the utilities produced by the other arms $\mathbf{u}=(u_{(\Tilde{m},\Tilde{M})})_{(\Tilde{m},\Tilde{M})\in\mathbb{A}}$ by repeating lines 4-9.
\STATE Update the weight of arm $(\Tilde{m},\Tilde{M}),\forall (\Tilde{m},\Tilde{M})\in \mathbb{A}$:
$$w_{(\Tilde{m},\Tilde{M})} \leftarrow w_{(\Tilde{m},\Tilde{M})} \exp\bigg(\frac{\eta(u_{(\Tilde{m},\Tilde{M})}-u^*)}{u^*}\bigg),$$where $u^*=\max_{(\Tilde{m},\Tilde{M})\in \mathbb{A}}u_{(\Tilde{m},\Tilde{M})}$.
\STATE Update the probability of choosing arm $(\Tilde{m},\Tilde{M})$: $$p_{(\Tilde{m},\Tilde{M})} = w_{(\Tilde{m},\Tilde{M})}/\sum_{(\Tilde{m},\Tilde{M})\in\mathbb{A}}w_{(\Tilde{m},\Tilde{M})}.$$
\ENDFOR
\end{algorithmic}
\end{algorithm}

\begin{figure}
    \centering
    \includegraphics[width=0.95\linewidth]{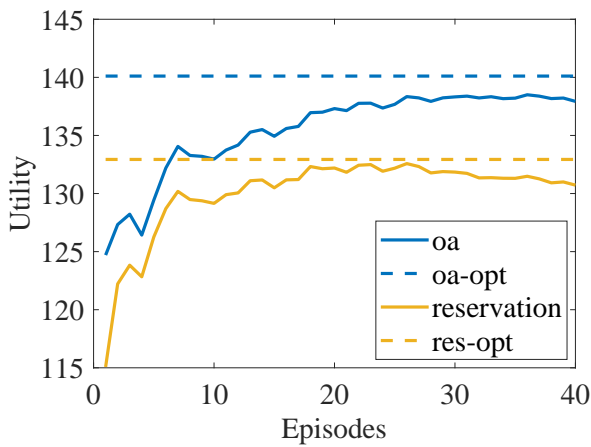}
    \caption{Convergence of the learning algorithm. The $y$-axis is the average value of 100 independent runs of utility produced by Algorithm \ref{alg:exp-weight}. \textbf{Alg-opt} is the utility of the algorithm with the best set of parameters. \textbf{Alg} is the performance of the algorithm with learned parameters. The arrivals generating this figure are from the truncated Gaussian distribution with mean = $\frac{m+M}{2}$, var = 1.}
    \label{fig:converge-learning}
\end{figure}

For a fair comparison, we also implement the adaptive version for the reservation-based algorithm. Fig. \ref{fig:adp-avr-case} and Fig. \ref{fig:adp-ext-case} show the utilities of our algorithm and its adaptive version, the utilities of reservation-based algorithm and its adaptive version, under the average cases and the extreme cases. We see that the adaptive implementation always improves the performance of both Algorithm \ref{alg:oa} and the reservation-based compared to the fixed parameter chosen in Section \ref{subsec:plain}. It is due to that in the adaptive implementation, the parameters are dynamically chosen in accordance with the input pattern. Fig. \ref{fig:converge-learning} shows the convergence of the average utility produced by Algorithm \ref{alg:exp-weight} to that produced by the static optimal parameter pair with the increase of episodes. Compared with the previous method of choosing the parameter offline based on past traces, the online learning algorithm does not need prior information on the input or past history, and thus can be applied in wider application scenarios.

\section{Conclusions and Future Directions} \label{sec:conclusions}
% Conclusions
In this paper, we consider the online network utility maximization problem, develop an algorithm that makes the allocation based on the utilization levels, and give a tight competitive analysis of the proposed algorithm. We find that the competitive ratio is linear in the number of links in the network. Extensive trace-driven simulations are conducted to show the empirical performance of the proposed algorithm. We confirm that the algorithm suffers from the common weakness for algorithms designed with the worst-case performance guarantee, whose performance under typical cases is usually mediocre. To further improve the performance, we first demonstrate the performance improvement when the algorithm parameters are chosen offline. Then we cast the parameter selection process as an online learning problem and show the performance gain by applying an off-the-shelf learning algorithm.

The ONUM problem is a relatively new problem for online algorithm design. We list some of the promising future directions that could be explored. First, we consider homogeneous capacities in this paper for the technical simplicity. It is more practical and intriguing to deal with heterogeneous capacities. We speculate that it is doable by a more fine-grained analysis. Second, the competitive ratio of our proposed algorithm is linear in terms of the number of links, but we make the conjecture that it is possible to further improve the competitive ratio by redesigning the threshold function and making more assumptions, such as utilizing the estimate of the demand information in \cite{zhang2017optimal}. Third, it is interesting to look for other real-life applications that can be modeled by the ONUM.

\bibliographystyle{IEEEtran}
\bibliography{main}

% \appendix
% \input{sections/appendix}

\end{document}